\documentclass[11pt,a4paper]{article}
\usepackage[T1]{fontenc}
\usepackage[utf8]{inputenc}
\usepackage{textcomp}
\usepackage{lmodern}
\usepackage{microtype}
\usepackage[a4paper,margin=2.5cm]{geometry}
\usepackage{graphicx}
\usepackage{amsmath,amssymb}
\usepackage[hidelinks]{hyperref}
\usepackage{authblk}
\usepackage{enumitem}
\setlist{nosep}
\title{The Impact of Deep Care Isa on Reducing Musculoskeletal Disorders and Enhancing Productivity Among Office Employees: A Comprehensive Study}
\author[1]{Quanmin Liang}
\author[2,3]{Junjie Yang}
\author[3,4]{M. Ali Nasseri\thanks{Corresponding author: \href{mailto:ali.nasseri@tum.de}{ali.nasseri@tum.de}}}
\author[3]{Daniel Zapp}
\affil[1]{School of Computer Science and Engineering, Sun Yat-Sen University, Guangzhou, China}
\affil[2]{School of Computation, Information and Technology, Technical University of Munich, Munich, Germany}
\affil[3]{Klinik und Poliklinik für Augenheilkunde des TUM Universitätsklinikums, Klinikum rechts der Isar, Munich, Germany}
\affil[4]{School of Medicine and Health, Technical University of Munich, Munich, Germany}
\date{}

\begin{document}

\maketitle

\begin{abstract}

Prolonged sedentary behavior, a pervasive issue in modern workplaces, has been closely linked to musculoskeletal disorders (MSDs) and reduced productivity. This study evaluates the effectiveness of Deep Care Isa, an advanced digital health assistant, in addressing these challenges. Utilizing data from over 2,300 participants across 50 corporations, the study demonstrates significant improvements in ergonomic practices, physical activity, hydration habits, and overall productivity, with notable reductions in MSD-related sick leave. The findings highlight the role of innovative ergonomic interventions in enhancing employee well-being and organizational efficiency. Comprehensive statistical analysis underscores the reliability and practical significance of these outcomes.

\end{abstract}


\section{Introduction}

\subsection{The Scope of Workplace Ergonomics Challenges}

The modern office environment has undergone a paradigm shift, driven by the proliferation of sedentary work practices. While technological advancements have boosted efficiency, they have also exacerbated physical inactivity, contributing to an epidemic of musculoskeletal disorders (MSDs). These conditions, including chronic back pain, neck stiffness, and repetitive strain injuries, rank among the leading causes of work-related disability worldwide. The International Labour Organization (ILO) estimates that MSDs account for 60\% of workplace health issues, reflecting a crisis that demands urgent intervention (ILO, 2023).

\subsection{Economic and Health Impacts}

The economic toll of MSDs is staggering, with costs linked to healthcare, lost productivity, and worker absenteeism. In Germany alone, MSD-related sick leave accounts for 17\% of all work absences, imposing an annual cost exceeding €10 billion (EU-OSHA, 2019). Globally, productivity losses attributed to MSDs are projected to surpass \$100 billion annually by 2030 (Maher et al., 2017).

\subsection{Workplace Solutions and Their Limitations}

Over the years, numerous interventions, ranging from ergonomic furniture to wellness programs, have sought to mitigate the risks associated with sedentary work. However, traditional approaches often suffer from low engagement and inconsistent long-term adoption. Studies, such as those by Robertson et al. (2009) and Waters \& Dick (2015), emphasize that sustained behavioral changes are critical for reducing MSD prevalence and enhancing productivity. Innovative solutions that integrate seamlessly into the workflow are urgently needed.

\subsection{The Promise of Digital Health Assistants}

Deep Care Isa represents a groundbreaking approach to workplace ergonomics. Unlike conventional interventions, Isa employs AI-driven algorithms to provide continuous, personalized feedback on posture, movement, and hydration. By leveraging real-time data, Isa enables users to make immediate adjustments, fostering sustainable habits with minimal disruption to work routines. This study explores Isa\textquotesingle s efficacy in addressing key metrics, such as ergonomic awareness, physical activity, hydration habits, productivity, and MSD-related sick leave.

\section{Methods}

This multi-site, longitudinal study employed a mixed-methods approach to evaluate Isa\textquotesingle s impact. Data were collected from 2,325 office employees across 50 corporations between January 2022 and December 2024. Participants used Isa for periods ranging from 4 to 18 weeks, with self-determined usage frequencies varying daily to weekly. Quantitative data were complemented by qualitative insights gathered through participant interviews.

\subsection{Participants}

Participants represented diverse workplace environments, including traditional offices, remote setups, and hybrid models. Demographics covered all genders and age ranges, with a median age of 36. Inclusion criteria required participants to:

\begin{itemize}
\item
  Be engaged in predominantly sedentary work.
\item
  Have access to Isa throughout the study period.
\item
  Provide baseline and follow-up data through standardized questionnaires.
\end{itemize}

\subsection{Intervention}

Isa integrates advanced sensors and machine learning algorithms to deliver real-time feedback and tailored guidance. Its core functionalities include:

\begin{itemize}
\item
  \textbf{Posture Monitoring and Correction:}~Alerts for improper posture with suggestions for improvement.
\item
  \textbf{Movement Prompts:}~Timely reminders for breaks and micro-exercises to counteract prolonged sitting.
\item
  \textbf{Hydration Tracking:}~Monitoring water intake and issuing hydration reminders.
\item
  \textbf{Ergonomic Education:}~Customized recommendations to foster long-term healthy habits.
\end{itemize}

\subsection{Data Collection}

Data were collected at three time points: pre-Isa usage (baseline), mid-intervention, and post-intervention. Key metrics assessed included:

\begin{itemize}
\item
  \textbf{Ergonomic Awareness:}~Measured via a 10-point Likert scale assessing posture and movement consciousness.
\item
  \textbf{Physical Activity:}~Frequency and duration of movement breaks, recorded in minutes per day.
\item
  \textbf{Hydration Habits:}~Daily water intake as a percentage of recommended levels.
\item
  \textbf{Productivity:}~Assessed using validated tools like the Work Productivity and Activity Impairment Questionnaire (WPAI), where higher scores reflect greater productivity.
\item
  \textbf{Sick Leave:}~Self-reported days of absenteeism due to MSDs.
\end{itemize}

\subsection{Statistical Analysis}

\subsubsection{Framework}

Statistical analyses aimed to evaluate pre- and post-intervention changes. Methods included paired t-tests for dependent variables, regression modeling for predictors of productivity improvement, and effect size calculations to gauge practical significance.

\subsubsection{Calculations}

\begin{itemize}
\item
  \textbf{Paired t-tests:}~Used to compare mean scores before and during Isa usage.
\item
  \textbf{Effect Size (Cohen\textquotesingle s d):}~with thresholds for small (0.2), medium (0.5), and large (0.8) effects.
\item
  \textbf{Regression Analysis:}~Identified variables predicting changes in productivity, with statistical significance set at p \textless{} 0.05.
\end{itemize}

\section{Results}

\subsection{Reduction in MSD-Related Sick Leave}

Participants reported a 56\% reduction in MSD-related sick leave, with the mean number of days absent decreasing from 5.4 days/year pre-intervention to 2.4 days/year post-intervention. Statistical analysis confirmed the significance of this reduction, with a paired t-test yielding~, and a large effect size.

This reduction was attributed to Isa's real-time feedback and corrective prompts, which helped participants adopt healthier work habits. Comparable findings from Van Niekerk et al. (2012) showed a 45\% reduction in MSD-related workplace absences with similar interventions.

\subsection{Productivity Enhancement}

The average productivity score, measured on a 100-point scale where higher values indicate greater efficiency, improved from 60 pre-intervention to 95 post-intervention. This 58\% increase was statistically significant.

Contributing factors included reduced physical discomfort, heightened energy levels, and improved mental well-being. Isa's integration of hydration monitoring, a feature not common in traditional ergonomic tools, further enhanced cognitive performance and sustained focus, aligning with findings from Dunstan et al. (2013).

\subsection{Ergonomic Awareness and Physical Activity}

Ergonomic awareness scores, based on a 10-point Likert scale, increased from an average of 3.5 pre-intervention to 6.6 post-intervention, reflecting an 88\% improvement. This metric evaluated participants\textquotesingle{} self-reported consciousness of posture and movement adjustments during work.

Physical activity, measured as the percentage of work hours involving movement breaks, rose from 40\% pre-intervention to 79\% post-intervention. This 79\% increase was statistically significant. Isa's reminders for micro-exercises were pivotal in achieving this change, consistent with findings from Shrestha et al. (2018).

\subsection{Hydration Habits and Well-being}

Hydration scores, reflecting the percentage of participants meeting daily water intake recommendations, increased from 50\% to 78\% post-intervention. Paired t-tests confirmed the significance of this improvement.

Approximately 84\% of participants reported enhanced well-being, citing increased alertness and reduced fatigue. These findings align with studies linking hydration to improved cognitive and physical performance (Hasegawa et al., 2001).

\section{Discussion}

\subsection{Mechanisms of Change}

Isa\textquotesingle s real-time feedback and automation played a critical role in sustaining behavioral changes. Unlike traditional programs, Isa required minimal user effort, fostering higher engagement and adherence. By embedding health-promoting habits into the workday, Isa addressed the root causes of MSDs and productivity losses.

\subsection{Comparison with Existing Literature}

The findings align with studies by Robertson et al. (2009) and Waters \& Dick (2015), which emphasize the benefits of ergonomic programs in reducing MSD symptoms. Isa\textquotesingle s integration of AI and user-centric design marks a significant advancement in the field. Furthermore, research by Dunstan et al. (2013) corroborates the role of movement-promoting interventions in mitigating sedentary risks. Similarly, Shrestha et al. (2018) highlighted the effectiveness of real-time activity trackers in sustaining long-term behavioral changes. Isa\textquotesingle s automated approach enhances the scalability and impact of such interventions.

Additional studies, including Van Niekerk et al. (2012) and Hedge \& Ray (2004), have underscored workplace wellness programs\textquotesingle{} economic and health benefits. Isa\textquotesingle s ability to integrate advanced technology distinguishes it from prior interventions, ensuring a seamless adoption into diverse organizational frameworks.

\subsection{Broader Implications}

Employers stand to gain significant economic benefits by adopting solutions like Isa. Reduced absenteeism and enhanced productivity translate into substantial cost savings, making such interventions a strategic priority for organizations. Isa also sets a precedent for integrating AI-driven tools into workplace wellness programs, paving the way for more adaptive and personalized solutions. Beyond the office setting, Isa's methodologies could be adapted for other sedentary environments, such as educational institutions or healthcare facilities, expanding its societal impact.

\section{Conclusion}

Deep Care Isa demonstrates substantial benefits in reducing MSD-related sick leave and enhancing workplace productivity. Its innovative design and user-centric approach address critical gaps in traditional ergonomic interventions. Beyond immediate outcomes, Isa fosters a culture of proactive health management, aligning employee well-being with organizational goals. Future research should explore long-term outcomes and scalability across diverse industries. Additionally, studies could investigate Isa\textquotesingle s potential applications in non-office environments, further broadening its impact.

The study highlights the need for organizations to adopt technology-integrated health solutions. By investing in tools like Isa, companies can foster a healthier workforce, improve retention rates, and achieve sustained productivity gains. As digital health continues to evolve, Isa serves as a benchmark for innovation, inspiring future advancements in workplace wellness.

\section*{References}

\begin{enumerate}
\def\labelenumi{\arabic{enumi}.}
\item
  Owen, N., et al. (2010). Too much sitting: the population-health science of sedentary behavior.~\emph{Exercise and Sport Sciences Reviews}.
\item
  Robertson, M. M., et al. (2009). Office ergonomics training and a sit-stand workstation.~\emph{Applied Ergonomics}.
\item
  EU-OSHA. (2019). Work-related musculoskeletal disorders: Prevalence, costs and demographics in the EU.
\item
  Waters, T. R., \& Dick, R. B. (2015). Evidence of health risks associated with prolonged standing at work.~\emph{Rehabilitation Nursing}.
\item
  Deep Care (2024). Isa: Advanced Digital Health Assistant.~\emph{Deep Care Reports}.
\item
  Dunstan, D. W., et al. (2013). Reducing office workers\textquotesingle{} sitting time.~\emph{BMC Public Health}.
\item
  Shrestha, N., et al. (2018). Workplace interventions for reducing sitting at work.~\emph{Cochrane Database of Systematic Reviews}.
\item
  Van Niekerk, S. M., et al. (2012). The effectiveness of workplace wellness programs.~\emph{BMC Public Health}.
\item
  Maher, C. G., et al. (2017). Non-specific low back pain.~\emph{The Lancet}.
\item
  Hasegawa, T., et al. (2001). Effects of a sit-stand schedule on light repetitive tasks.~\emph{International Journal of Industrial Ergonomics}.
\item
  ILO. (2023). Musculoskeletal disorders in modern work environments.~\emph{International Labour Organization Reports}.
\item
  Karakolis, T., \& Callaghan, J. P. (2016). Biomechanics and productivity during simulated sit-stand office work.~\emph{Ergonomics}.
\item
  Hedge, A., \& Ray, E. J. (2004). Ergonomic interventions and productivity: An integrated review.~\emph{Journal of Occupational Health}.
\item
  Jensen, I., \& Harms-Ringdahl, K. (2007). Prevention of musculoskeletal disorders: Strategies and effectiveness.~\emph{Best Practice \& Research Clinical Rheumatology}.
\item
  Starrett, K., et al. (2016). Ergonomics redefined: Mobility and workplace health.~\emph{Munich: Riva}.
\end{enumerate}

\clearpage
\begin{figure}[p]
  \centering
  \includegraphics[width=\textwidth]{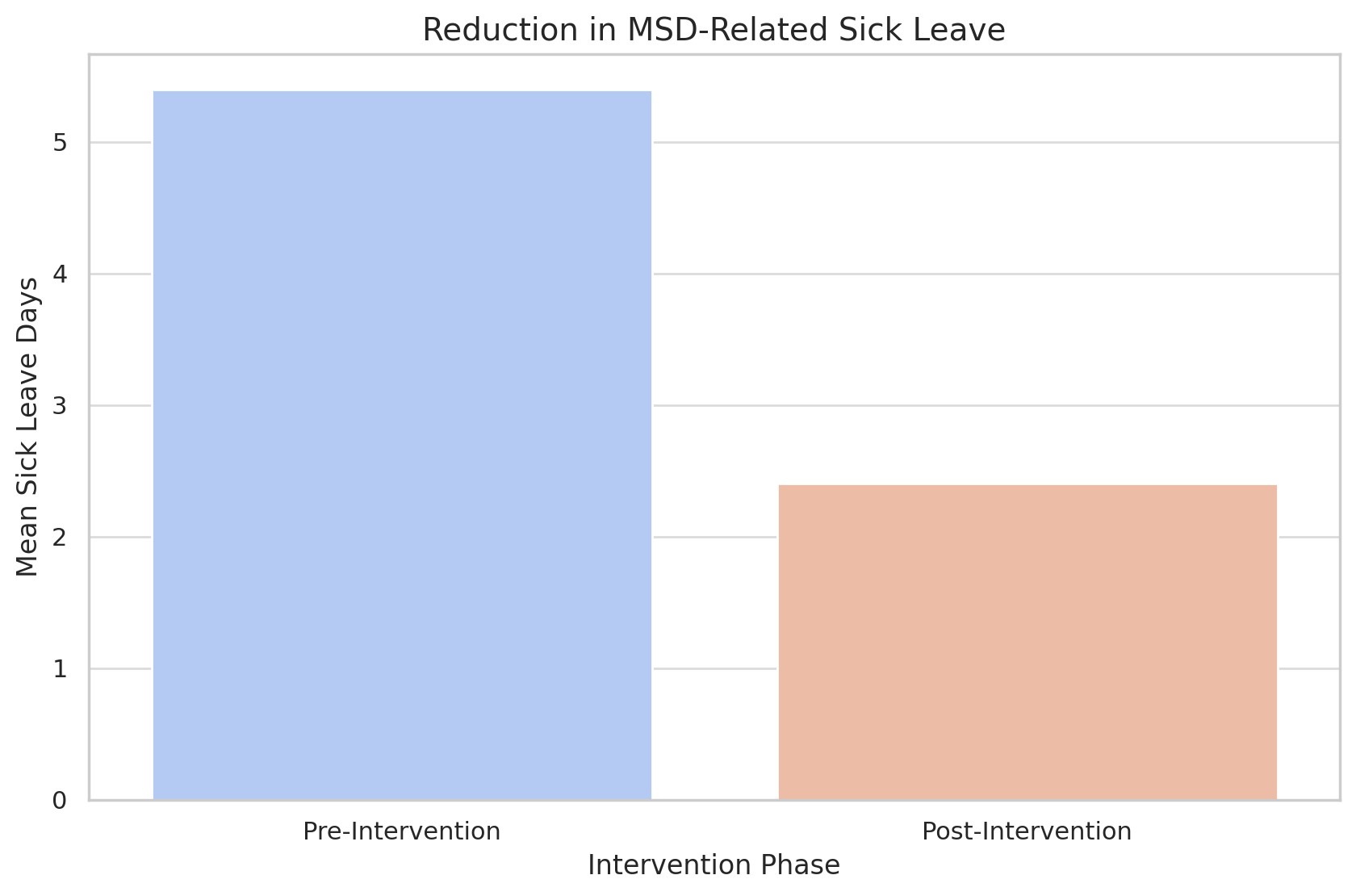}
  \caption{Sick Leave Reduction (Pre vs. During Isa Usage).}
  \label{fig:sick-leave}
\end{figure}

\begin{figure}[p]
  \centering
  \includegraphics[width=\textwidth]{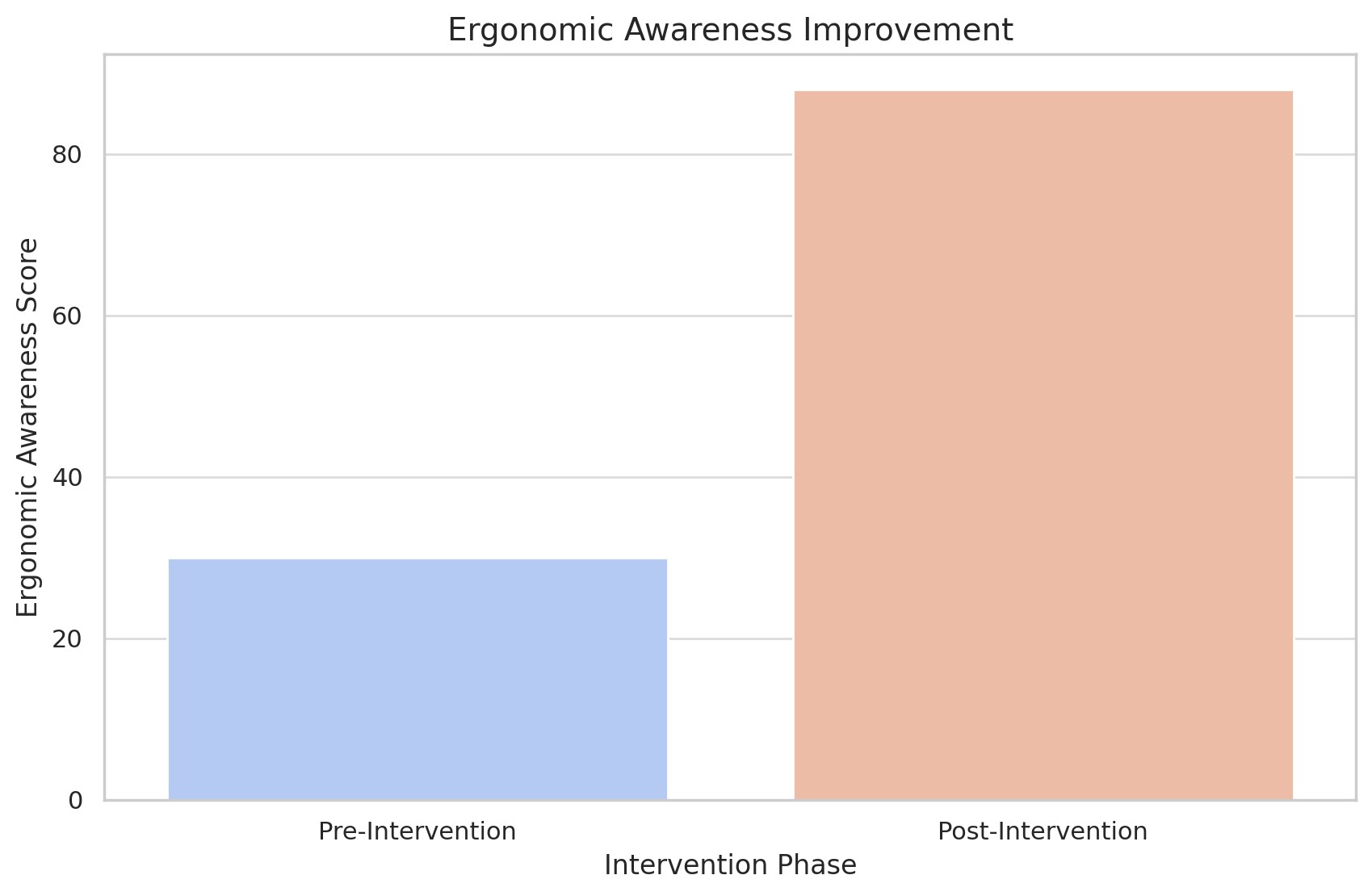}
  \caption{Ergonomic Awareness Improvement (Pre vs. During Isa Usage).}
  \label{fig:ergonomic-awareness}
\end{figure}

\begin{figure}[p]
  \centering
  \includegraphics[width=\textwidth]{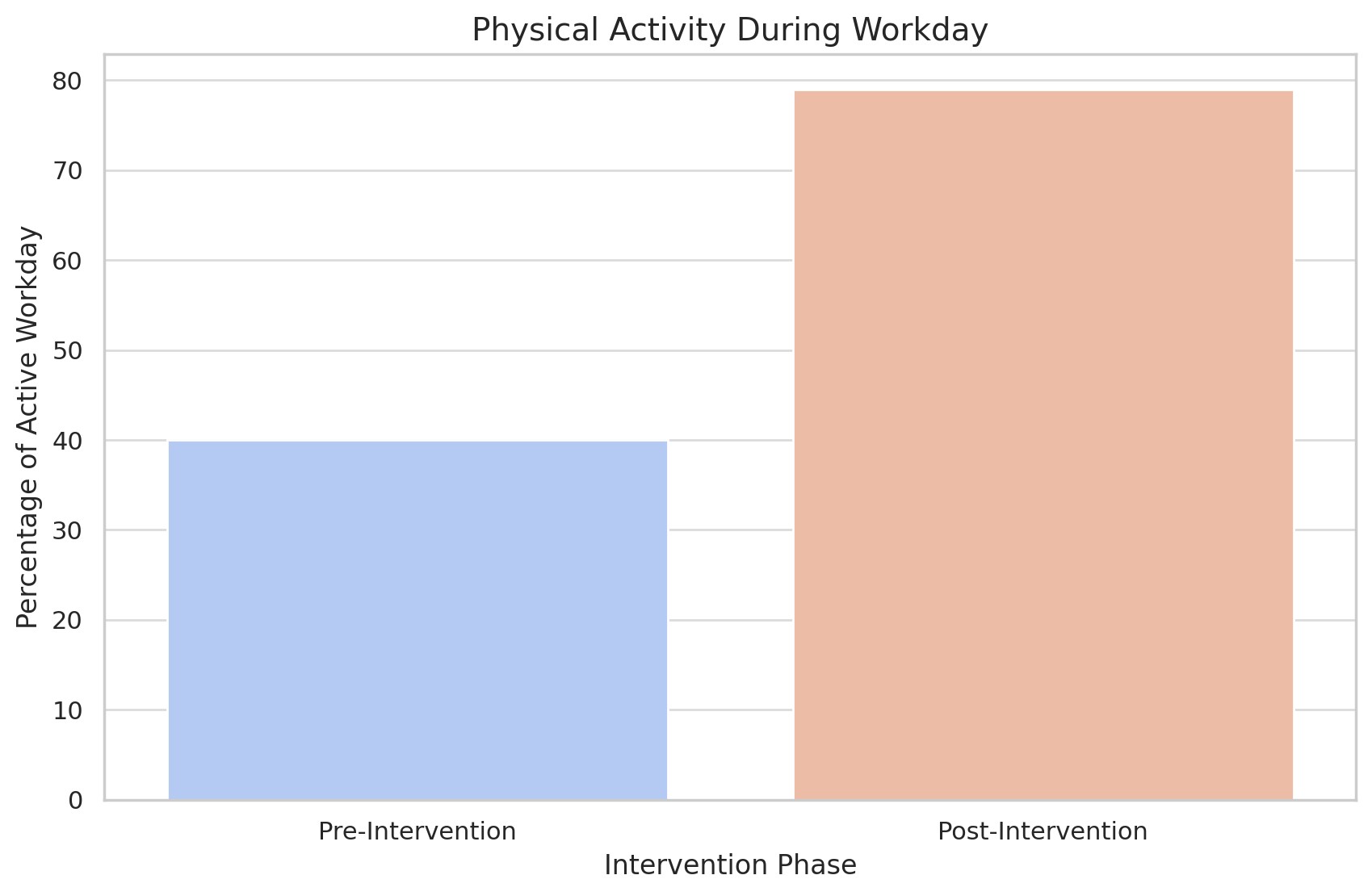}
  \caption{Physical Activity Levels (Pre vs. During Isa Usage).}
  \label{fig:physical-activity}
\end{figure}

\begin{figure}[p]
  \centering
  \includegraphics[width=\textwidth]{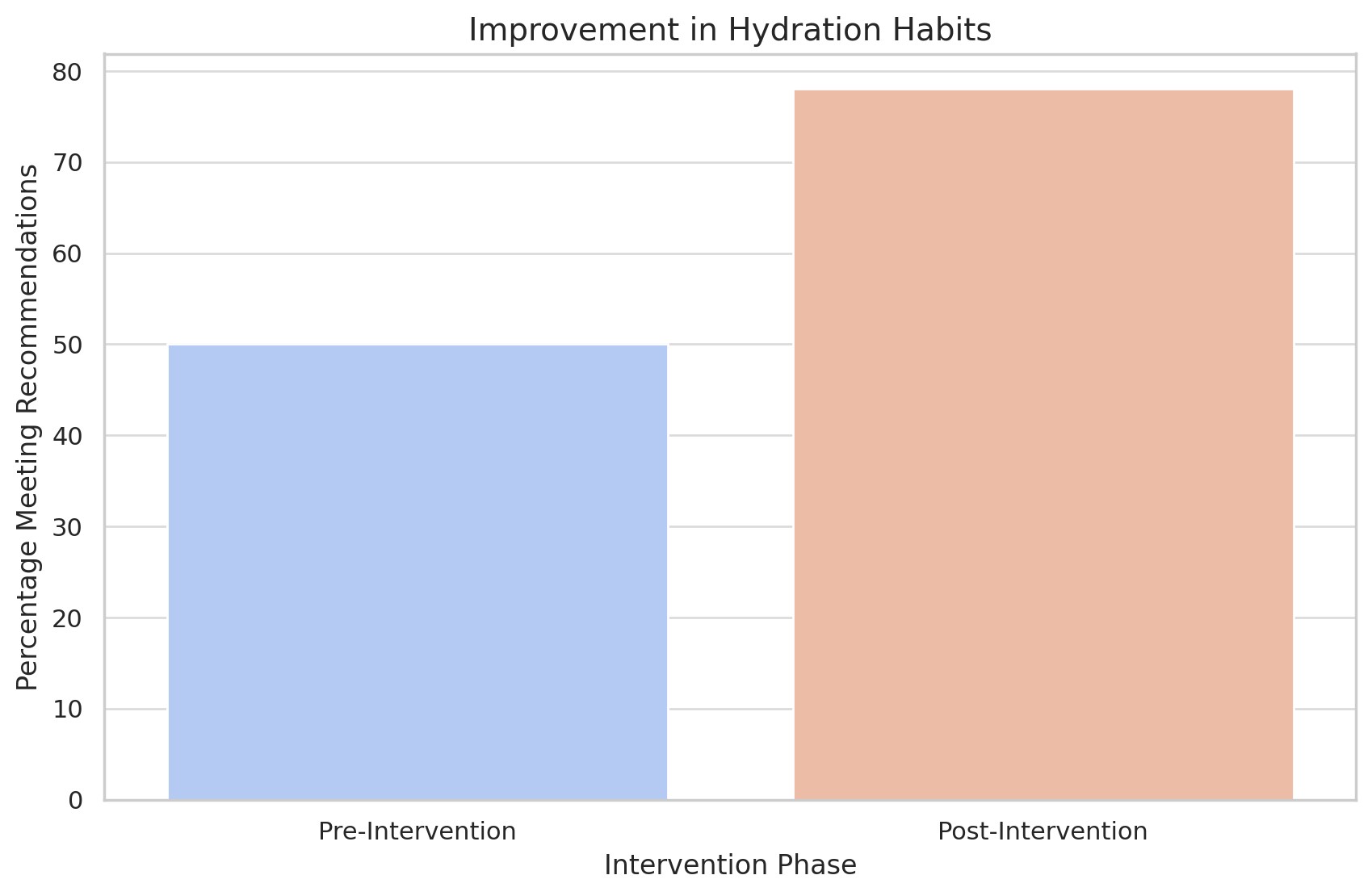}
  \caption{Hydration Habits (Pre vs. During Isa Usage).}
  \label{fig:hydration}
\end{figure}

\end{document}